# Role of seed layer in growing atomically flat $TiTe_2/Sb_2Te_3$ heterostructure thin films at the wafer scale

Chao Nie[1], Xueyang Shen[1], Junying Zhang[1], Chenyu Wen[1], Yuxin Du[1], Yazhi Xu[2], En Ma[1], Xiaozhe Wang[1,*], Riccardo Mazzarello[3,*], Wei Zhang[1,*], Jiang-Jing Wang[1,*]

[1]Center for Alloy Innovation and Design (CAID), State Key Laboratory for Mechanical Behavior of Materials, Xi'an Jiaotong University, Xi'an 710049, China.
[2]Department of Applied Physics, Chang'an University, Xi'an 710049, China.
[3]Department of Physics, Sapienza University of Rome, Rome 00185, Italy.

Emails: wangxiaozhe@xjtu.edu.cn, riccardo.mazzarello@uniroma1.it, wzhang0@mail.xjtu.edu.cn, j.wang@xjtu.edu.cn

**Abstract:**
Chalcogenide phase-change materials (PCMs) are a leading candidate for advanced memory and computing applications. Epitaxial-like growth of chalcogenide thin films at the wafer scale is important to guarantee the homogeneity of the thin film but is challenging with magnetron sputtering, particularly for the growth of phase-change heterostructure (PCH), such as $TiTe_2/Sb_2Te_3$. In this work, we report how to obtain highly textured $TiTe_2/Sb_2Te_3$ heterostructure thin films with atomically sharp interfaces on standard silicon substrates. By combining atomic-scale characterization and *ab initio* simulations, we reveal the critical role of the $Sb_2Te_3$ seed layer in forming a continuous Si-Sb-Te mixed transition layer, which provides a wafer-scale flat surface for the subsequent epitaxial-like growth of $TiTe_2/Sb_2Te_3$ thin film. By gradually reducing the thickness of the seed layer, we determine its critical limit to be ~2 nm. Non-negligible in-plane tensile strain was observed in the $TiTe_2$ slabs due to the lattice mismatch with the adjacent $Sb_2Te_3$ ones, suggesting that the chemical interaction across the structural gaps in the heterostructure is stronger than a pure van der Waals interaction. Finally, we outline the potential choices of chalcogenides for atomically flat seed layers on standard silicon substrates, which can be used for wafer-scale synthesis of other high-quality PCM or PCH thin films.

**Keywords:** phase-change heterostructure, seed layer, thin film, atomic characterizations, $TiTe_2/Sb_2Te_3$

## 1. Introduction:

Chalcogenide phase-change materials (PCMs) are one of the leading candidates for non-volatile memory and neuromorphic in-memory computing applications [1-12]. The most important family of PCM consists of the Ge-Sb-Te alloys along the GeTe-$Sb_2Te_3$ pseudo-binary line [13-21], e.g. $Ge_2Sb_2Te_5$ (GST), which has been extensively used in electronic, photonic as well as hybrid optoelectronic devices [22-30]. The basic working principle is the large contrast in electrical resistance or optical transmission between the crystalline and amorphous phases of PCM for memory encoding [1]. Given its metastable nature, the amorphous phase undergoes spontaneous structural relaxation, causing a steady increase in electrical resistance with time, known as the resistance drift [31-34]. For binary storage, this aging effect is not a serious issue, as the contrast window is enlarged instead of reduced with time. But for the multilevel programming scheme adopted in neuromorphic devices, the drift issue is a major hindrance for high-precision encoding of multiple resistance states [35-37].

The structural origin of the resistance drift in amorphous PCMs stems from the disappearance of structural defects that are generated during the rapid cooling process and the reinforcement of Peierls-like distortions between pairs of octahedral bonds with time [38-41]. Scaling down the film thickness to exploit volume effects [42-45] or alloying PCM with impurities can help reduce the drift coefficient effectively [46-49]. Another promising approach to suppress drift is based on heterostructures composed of alternating thin layers of two different PCMs or of a PCM and a confinement material (CM). Such heterostructures are typically grown by heterogeneous sputtering or molecular beam epitaxy. Besides the low drift, they also offer low switching energy and fast switching speed. Prominent examples include the GeTe(GST)/$Sb_2Te_3$ [50-59] superlattice heterostructures, which are made of two standard PCMs with similar melting point, and the $TiTe_2$/$Sb_2Te_3$ phase-change heterostructures (PCH) [60-65]. In the latter, $TiTe_2$ serves as CM due to its much higher melting point than $Sb_2Te_3$, effectively reducing the drift coefficient of the PCM layers. More heterostructure combinations were predicted [66] and developed [67-70] for memory applications with CM = $TiTe_2$, $MoTe_2$, $NiTe_2$ or $HfTe_2$, and PCM = GeTe, GST, $Ge_4Sb_6Te_7$ or $Ta_xSb_2Te_3$.

For practical use, it is important to form sharp interfaces between the CM layers and PCM layers during growth. Otherwise, accumulation of structural defects at the interfaces upon extensive cycling could turn the PCM into conventional transition metal doped alloys, causing device degradation [60, 61]. In this work, we report how to deposit $TiTe_2$/$Sb_2Te_3$ heterostructure thin films with atomically sharp interfaces at the wafer scale. The key to obtaining high-quality heterostructure thin films relies on the quality of the seed layer deposition and the *in situ* heating during subsequent co-sputtering of the heterostructure. We determine the critical (minimum) thickness of the seed layer required to ensure the formation of sharp interfaces on a standard silicon substrate, and provide in-depth understanding of this interface region via atomic-scale structural characterization experiments and *ab initio* calculations. We also find that the $TiTe_2$ crystal lattice is under non-negligible tensile strain due to the lattice mismatch with crystalline $Sb_2Te_3$. At last, we provide guidelines on the choice of seed layer that can guarantee highly textured thin film deposition.

## 2. Experimental and simulation details

The $TiTe_2$/$Sb_2Te_3$ heterostructure thin films were deposited on silicon wafers by magnetron sputtering (AJA, Orion-8). The silicon wafers were cleaned by sonication in acetone, ethanol, and deionized water for 10 min, respectively, and then baked at 80°C for 30 minutes. The base pressure in the deposition chamber was maintained at ~1×$10^{-7}$ Torr. The $Sb_2Te_3$ layers were deposited using a $Sb_2Te_3$ (99.99%) target at a pressure of 3 mTorr, and the $TiTe_2$ layers were deposited by co-sputtering with Ti (99.995%) and Te (99.99%) targets at a pressure of 4.7 mTorr. The cross-section specimens of the $TiTe_2/Sb_2Te_3$ thin films were prepared using a FEI Helios NanoLab 600i focus ion beam (FIB) system with a Ga ion beam at 30 kV beam energy, and were thinned and polished at 5 kV / 20 pA and 2 kV / 10 pA, respectively. The spherical aberration corrected scanning transmission electron microscopy (STEM) high-angle annular dark-field (HAADF) and energy dispersive X-ray (EDX) mapping experiments were performed on a JEM-ARM300F2 STEM with a probe aberration corrector, operated at 300 kV. The high-resolution transmission electron microscopy (HRTEM) and EDX experiments were performed on a Talos-F200X operated at 200 kV. The atomic force microscopy (AFM) experiments were measured using a SPM-9700HT. The X-ray diffraction (XRD) experiments were performed by a Bruker D8 ADVANCE with CuKα radiation (λ = 1.54056) in the 2θ range 5°~60° with the scanning step size of 0.02°. The Raman spectra were collected by using a Renishaw inVia Qontor Raman microscope with a solid-state 532 nm laser for the excitation. The laser power was set to 0.25 mW, and the exposure time was 1 s with 100 cycles. Density functional theory (DFT) calculations were carried out using the Vienna Ab-initio Simulation Package (VASP) code [71]. The Perdew-Burke-Ernzerhof (PBE) functional [72], the projector augmented plane-wave (PAW) pseudopotentials [73], and the Grimme's D3 dispersion correction [74] were applied in the calculations. An energy cutoff of 500 eV was used. We constructed 2 × 2 × 5 $TiTe_2$ and 2 × 2 × 1 $Sb_2Te_3$ supercells that both contained 60 atoms, and used k-point mesh of 9 × 9 × 2 points used to sample the Brillouin zone. The Löwdin charges were calculated using the LOBSTER code [75-77].

## 3. Results and discussion

We deposited several $TiTe_2/Sb_2Te_3$ PCH thin films on 7 cm × 7 cm silicon wafers following the procedure shown in Fig. 1a. Before deposition, the silicon substrate was cleaned using Ar-plasma etching with radio frequency at ~35 Watt over 30 minutes in the high vacuum sputtering chamber. This cleaning process is needed to remove the native oxides on top of the substrate, and is frequently termed as the reverse sputtering process [78-80]. As soon as the etching was completed, a ~5 nm $Sb_2Te_3$ seed layer was deposited at room temperature. Next, the substrate was heated at ~280 °C for 60 min to fully crystallize the $Sb_2Te_3$ seed layer, forming a trigonal phase with ordered $Sb_2Te_3$ quintuple-layer (QL) blocks (each QL has a thickness of ~1 nm) [81-83]. Subsequently, a PCH unit made of a ~5 nm $TiTe_2$ slab, corresponding to ~7–8 $TiTe_2$ trilayer (TL) blocks, followed by a ~5 nm $Sb_2Te_3$ slab, were deposited via alternate sputtering of the $Sb_2Te_3$, Ti and Te targets under the *in situ* heating condition. In total, four PCH units were deposited. As the last step, the PCH thin film with a thickness of ~45 nm (including the seed layer) was annealed at 280 °C for another 30 min, and was naturally cooled to room temperature in the high vacuum chamber. The whole thin film was covered with a ~10 nm-thick ZnS-$SiO_2$ capping layer to prevent oxidation. The cross-sectional structural properties and elemental distributions of this $TiTe_2/Sb_2Te_3$ thin film were characterized by combining the HRTEM with EDX experiments. As shown in Fig. 1b, the HRTEM image provides an overview of the PCH thin film, which has well-aligned lattice fringes parallel to the substrate. In the corresponding EDX mapping, the Ti (red) and Sb (green) signals are well separated, while the Te (blue) signal is uniformly distributed throughout the PCH thin film. Only the Sb and Te signals were detected within the bottom ~5 nm region in contact with the substrate.

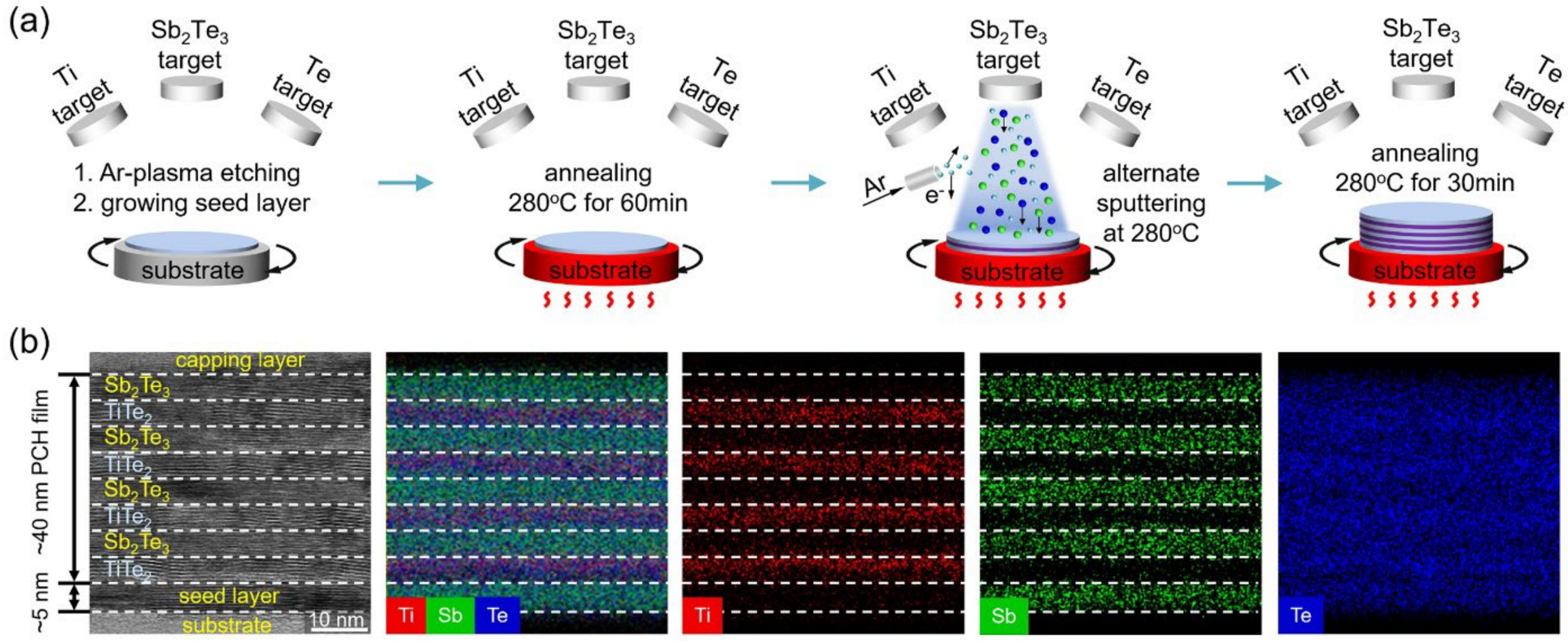

**Fig. 1 The deposition procedure and characterization of the $TiTe_2/Sb_2Te_3$ thin film.** (a) The schematic diagrams of the deposition process in four steps. (b) The HRTEM and EDX mapping images of the PCH thin film.

The epitaxial-like growth of the PCH thin film depends on the quality and out-of-plane texture of the $Sb_2Te_3$ seed layer. In Fig. 2, we characterized the surface morphology of a seed layer and a full PCH thin film. The optical images of the two thin films on the silicon wafers are presented in Fig. 2a and 2c. We selected 16 uniformly distributed regions across the two thin films and measured the surface roughness of each region by using AFM. For each AFM scan, the size of the measured region was set to 2 µm × 2 µm. Fig. 2b presents the AFM images of six regions of the seed-layer sample (the remaining 10 AFM images are included in Fig. S1a of the supplementary material). The respective root-mean-square roughness $R_q$ values provided in the figure range from ~4.41 Å to ~5.17 Å, indicating a highly smooth surface. The 16 AFM images of the PCH thin films are shown in Fig. 2d and Fig. S1b. The $R_q$ values range from ~6.50 Å to ~7.50 Å, which is slightly broader than that of the seed layer but is still smaller than the thickness of a single $Sb_2Te_3$ QL block. Therefore, we conclude that we have obtained a high-quality PCH thin film with smooth surface.

We carried out XRD and Raman spectroscopy experiments to investigate the structural properties of the obtained PCH thin films. We also prepared pristine $Sb_2Te_3$ and $TiTe_2$ thin films of ~45 nm thickness for comparison, for which a ~5 nm $Sb_2Te_3$ seed layer was used consistently. Fig. 3a and Fig. S2a show the measured XRD patterns at 16 locations of the PCH thin film. No obvious difference can be observed in these XRD curves. In comparison with the two pristine thin films, the PCH thin film exhibited a series of (00l)-oriented diffraction peaks, namely, the (003), (006), (009), (0015) and (0018) peaks of the $Sb_2Te_3$ slabs, and the (001) and (002) peaks of the $TiTe_2$ slabs. Note that the $Sb_2Te_3$ (009) and $TiTe_2$ (002) peaks overlap due to their close peak locations at 2θ ~26.3° and 27.0°, leading to a smaller shoulder in the primary peak of the PCH thin film. These well-defined (00l)-oriented diffraction patterns confirm the epitaxial-like growth with out-of-plane texture across the entire wafer.

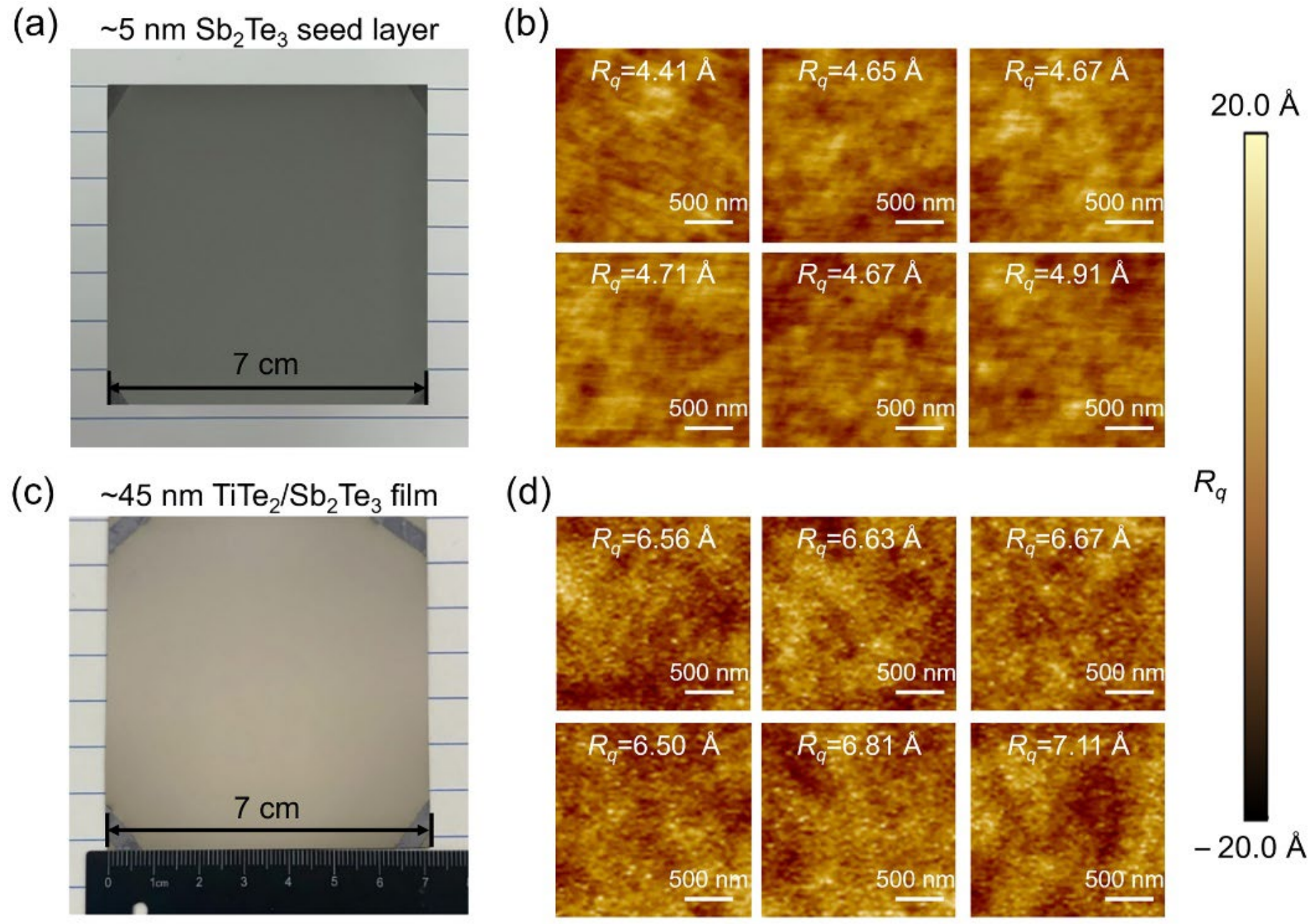

**Fig. 2 Surface morphology characterization of the seed layer and the PCH thin film.** (a) The optical image of the $Sb_2Te_3$ seed layer with a thickness of ~5 nm on a silicon wafer. (b) The AFM images taken at 6 different locations on the surface of the seed layer. (c) The optical image of the $TiTe_2/Sb_2Te_3$ thin film on a silicon wafer. (d) The AFM images taken at 6 different locations on the surface of the PCH thin film. $R_q$ is the measured root-mean-square roughness.

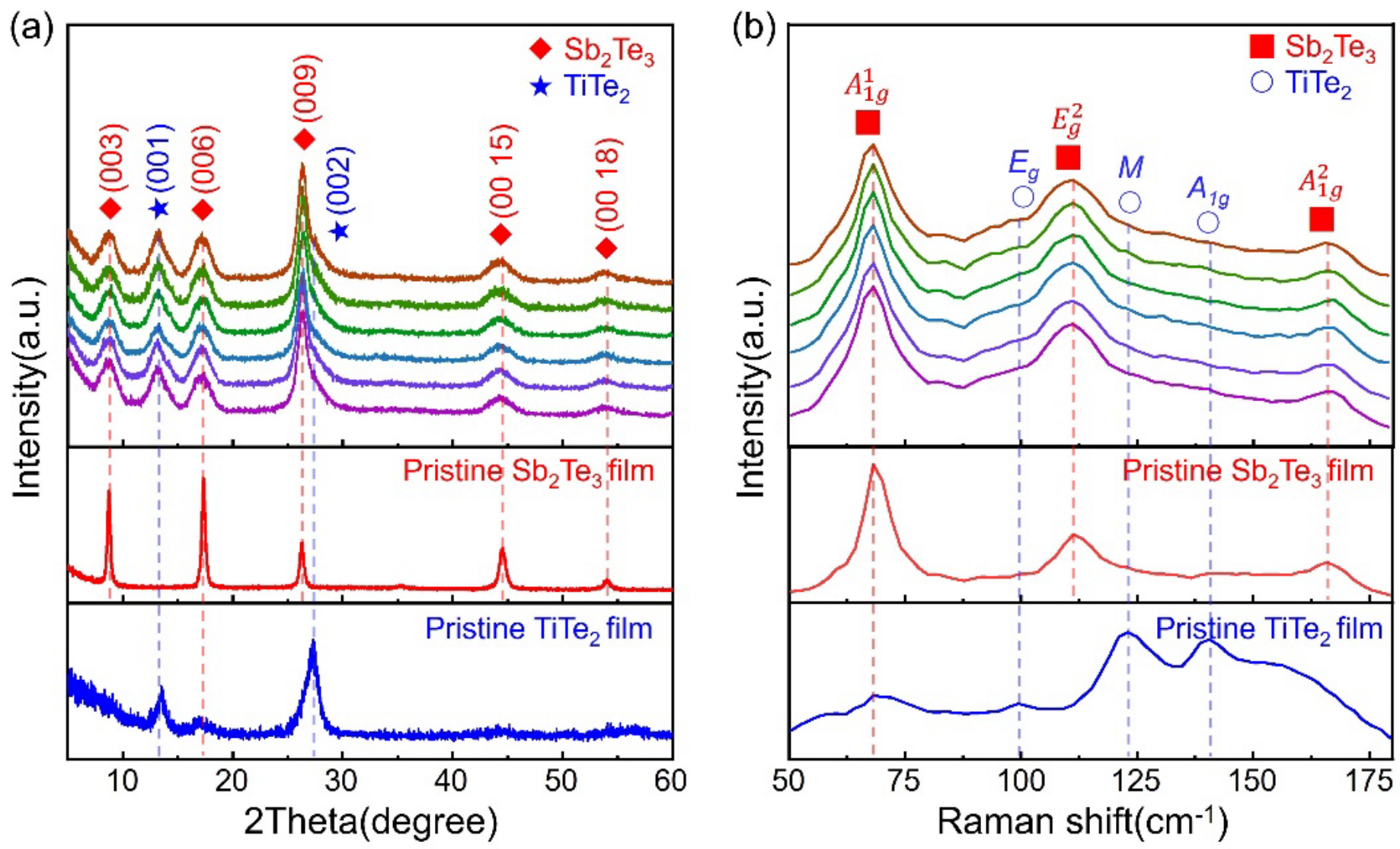

**Fig. 3 XRD and Raman characterizations.** The (a) XRD and (b) Raman characterizations carried out on various locations of the PCH thin film. The XRD and Raman curves of the pristine $Sb_2Te_3$ thin film and pristine $TiTe_2$ thin film are shown for comparison.

Fig. 3b and Fig. S2b present the Raman spectra measurements of the three thin films within a range of 50–180 $cm^{-1}$ using a 532 nm solid-state excitation laser. The measured local Raman spectra at 16 locations of the PCH thin film show three major peaks. By comparing with pristine $Sb_2Te_3$ [84-86], the three vibrational modes are identified as an out-of-plane mode $A_{1g}$ (1) at ~68.2 $cm^{-1}$, an in-plane mode $E_g$ (2) at ~111.3 $cm^{-1}$ and a second out-of-plane mode $A_{1g}$ (2) at ~165.8 $cm^{-1}$ for the PCH thin film. For the pristine $TiTe_2$ thin film, three major vibrational modes [87] were found at ~101.5 $cm^{-1}$ for $E_g$, ~124.9 $cm^{-1}$ for $M$ and ~142.4 $cm^{-1}$ for $A_{1g}$, respectively. However, these $TiTe_2$-related vibrational modes were not visible in the PCH thin film due to the much weaker Raman response. The highly consistently XRD patterns and Raman spectra across the entire $TiTe_2/Sb_2Te_3$ thin film are indicative of wafer-scale uniformity.

Next, we conducted STEM-HAADF imaging experiments to gain a better understanding of the atomic interfaces between $TiTe_2$ and $Sb_2Te_3$ in the heterostructure thin film. Fig. 4 presents the HAADF images of the pristine $Sb_2Te_3$, pristine $TiTe_2$ and $TiTe_2/Sb_2Te_3$ thin films in the $[01\bar{1}0]$ direction. The brightness of each spot is roughly proportional to $Z^2$, where Z represents the average atomic number of the atomic column along the incident direction of the electron beam [88]. The atomic layers in the $Sb_2Te_3$ blocks exhibit uniform image intensity, owing to the comparable atomic number of Sb (Z=51) and Te (Z=52). In contrast, the central atomic layer of the $TiTe_2$ blocks displays markedly reduced intensity due to the much smaller atomic number of Ti (Z=22). The size of the $Sb_2Te_3$ block in the pristine thin film and in the heterostructure thin film is comparable, but for the $TiTe_2$ blocks, the in-plane atomic spacing is enlarged from ~3.77 Å in the pristine thin film to ~3.86 Å in the PCH thin film. This in-plane expansion indicates that the lattice mismatch between $Sb_2Te_3$ and $TiTe_2$ induces a ~2.4% tensile strain in the $TiTe_2$ slabs of the PCH thin film. Similar behavior was observed in GeTe/$Sb_2Te_3$ thin films, where the tensile strain could reach ~1.6% in the GeTe slab as compared to bulk GeTe [89]. The observation of lattice expansion in the $TiTe_2$ slab suggests that the alternately stacked $Sb_2Te_3$ and $TiTe_2$ slabs are connected not only by pure van der Waals forces but also additional chemical interaction across the structural gaps. Therefore, the structural gaps in the $TiTe_2/Sb_2Te_3$ PCH should also be regarded as "vdW-like" gaps rather than pure vdW gaps, similar to those in GeTe/$Sb_2Te_3$ heterostructures, and those in trigonal bulk $Sb_2Te_3$ and GST [89-91]. In a recent work, we revealed the critical role of weak chemical interactions across the vdW-like gaps in stabilizing metavalent bonding (MVB) [92-95] in trigonal $Sb_2Te_3$ and GST, and explained how to tailor the degree of this in-gap interaction for altering the optical properties via uniaxial strain along the out-of-plane direction [96]. Although pristine $TiTe_2$ is regarded as a vdW material like other transition metal dichalcogenides [97], the inter-slab coupling can be significantly increased by co-sputtering with MVB-type layered solids, which induces in-plane strain affecting the electronic structure [98-100].

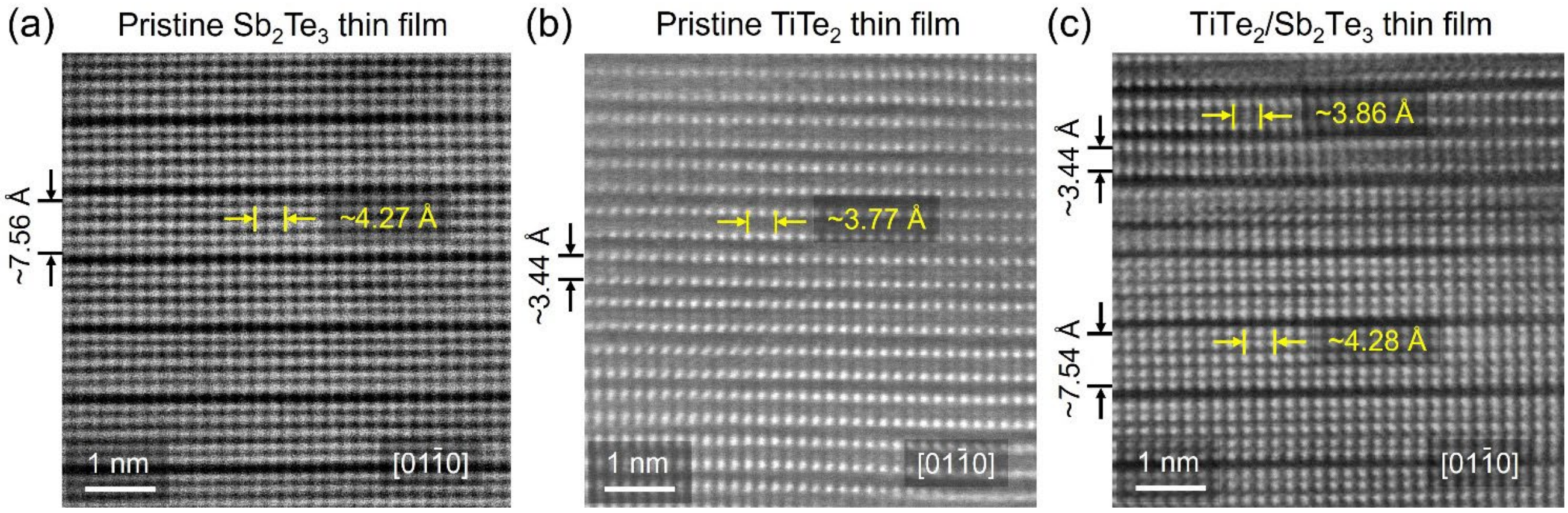

**Fig. 4 Atomic-scale characterization of the lattice parameters along the $[01\bar{1}0]$ direction.** The HAADF images of (a) the pristine $Sb_2Te_3$ thin film, (b) the pristine $TiTe_2$ thin film and (c) the $TiTe_2/Sb_2Te_3$ thin film.

The synthesis of the PCH thin film with wafer-scale uniformity is rendered possible by high-quality deposition of the seed layer. The HRTEM image in Fig. 5a shows an overall view of the multiple interfaces between different layers. We focus on the two interfaces indicated by the red and blue arrows, the corresponding atomic-scale HAADF images are shown in Fig. 5b and Fig. 5c. We used a standard silicon substrate with the <100> crystal orientation that was naturally covered with a ~4–5 nm $SiO_2$ layer. The initial Ar-plasma etching only removed native oxygen at the topmost of the substrate, not the full $SiO_2$ layer. The HAADF image (Fig. 5c) and the corresponding EDX maps (Fig. 5d–h) revealed a mixed Si-Sb-Te region of ~2 nm with no clear atomic columns except for the top three Sb-Te atomic layers. Above this transition region, we observed ordered $Sb_2Te_3$ QL blocks of the seed layer. Clearly, the ordered QL blocks are not directly contacted with the Si-100 surface. This transition region was formed because we conducted deposition of the $Sb_2Te_3$ seed layer immediately after Ar-plasma etching of the substrate. This procedure resulted in an amorphous Si layer of ~2–3 nm [78-80] with gradually reduced density from the substrate towards the seed layer (Fig. 5g), and the Si atoms intermixed with the subsequent sputtered Sb and Te atoms. It is important to note that the Ar-plasma etching process cannot be subjected to *in situ* heating, which markedly increases the probability of re-forming $SiO_2$. In such a case it is no longer feasible to obtain a flat seed layer for high-quality growth of PCH thin films (Fig. S3).

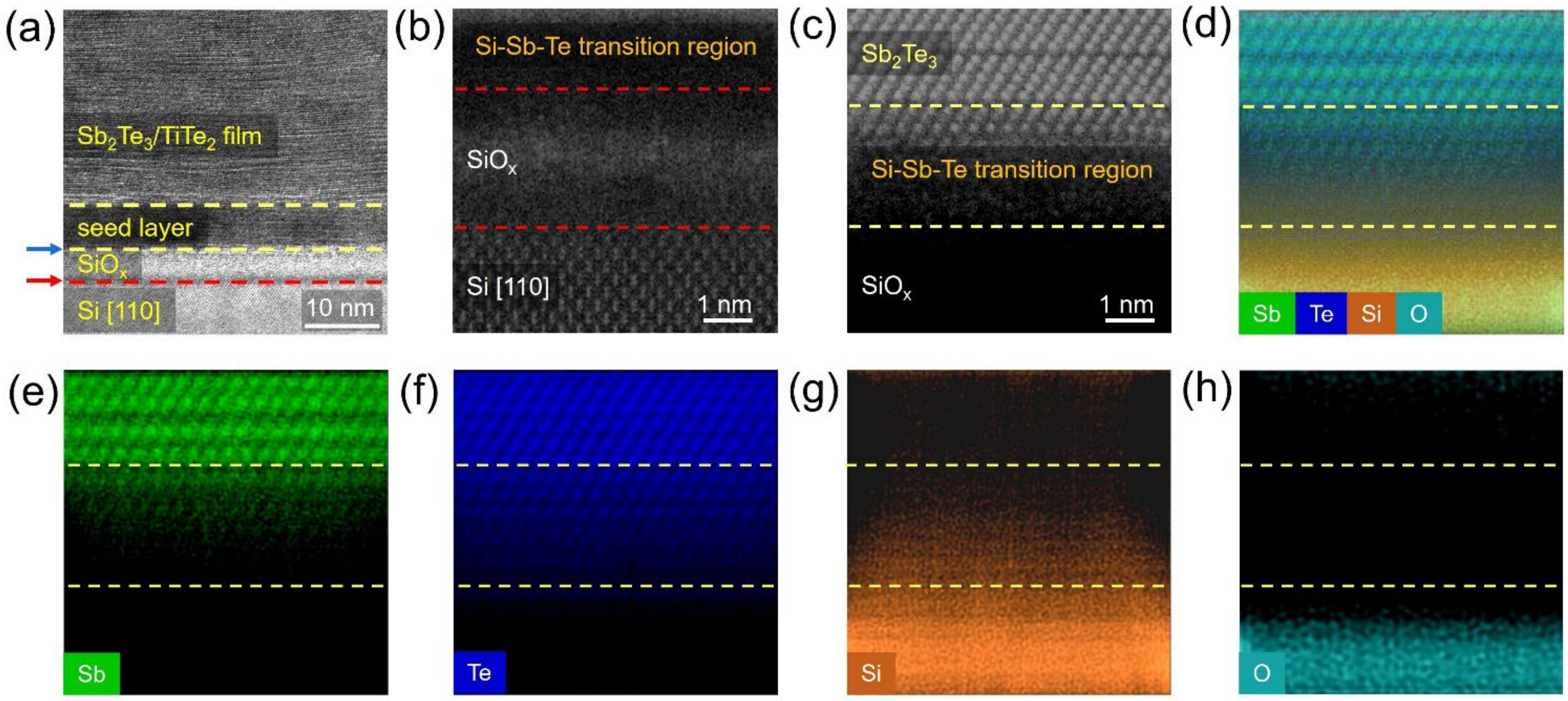

**Fig. 5 Structural and elemental characterizations of the $TiTe_2/Sb_2Te_3$ thin film.** (a) The large-scale HRTEM image of the multiple interfaces between the PCH thin film and the silicon substrate. (b) and (c) The atomic-resolution HAADF images of the interfaces marked by the red and yellow arrows in (a). (d)-(h) The corresponding EDX mapping of the transition region shown in (c).

The atomic-scale characterization of the seed layer suggests that its thickness $t_s$ ($Sb_2Te_3$) can be further optimized. We repeated the sputtering process shown in Fig. 1a but reduced $t_s$ ($Sb_2Te_3$) to ~2 nm, ~1 nm and finally 0 nm. In the latter case, the first ~5 nm $TiTe_2$ layer served as the seed layer. The HRTEM measurements showed that the PCH thin film with the ~2 nm $Sb_2Te_3$ seed layer exhibited well-aligned lattice fringes parallel to the substrate, as indicated by the white arrows in Fig. 6a. In the other two cases, the lattice fringes of the PCH thin films were no longer flat and smooth, as shown in Fig. 6b and Fig. 6c. The atomic-scale HAADF images of the seed layer region are shown in Fig. 6d-f for the three

PCH thin films. The first PCH thin film showed a similar Si-Sb-Te transition region with top 2–3 ordered Sb or Te atomic layers and a full $Sb_2Te_3$ QL block, consistent with that in the PCH thin film with $t_s$ ($Sb_2Te_3$) ~5 nm. However, the transition region in the other two PCH thin films failed to promote the formation of well-ordered atomic layers at the top edge. Therefore, the critical minimum thickness of the seed layer is ~2 nm for $Sb_2Te_3$ to generate a flat surface, i.e., a full intact $Sb_2Te_3$ QL block, for the subsequent growth of high-textured thin films on top. However, it is less suitable to use $TiTe_2$ as the seed layer, as sketched in Fig. 6 g-i.

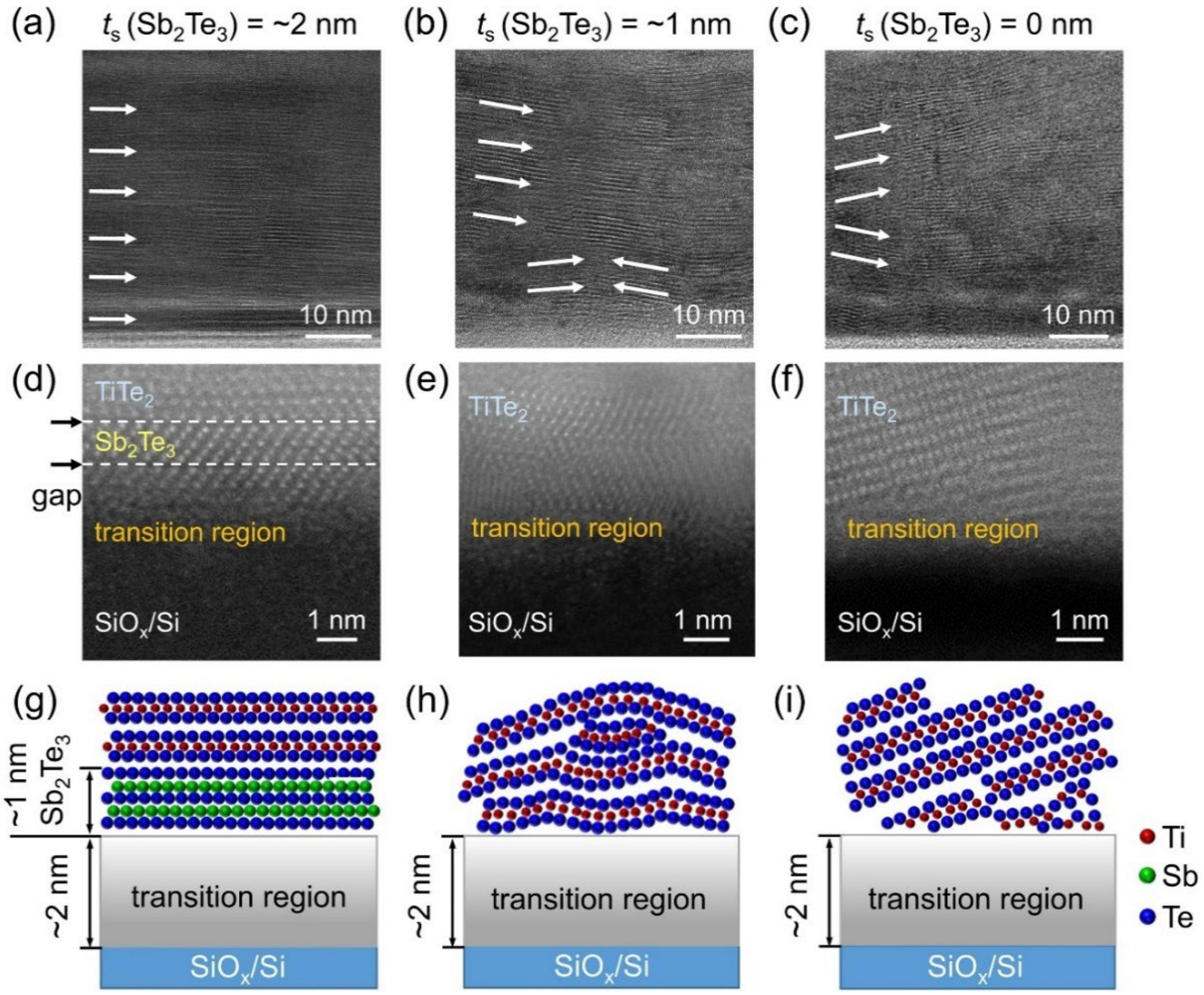

**Fig. 6 The critical thickness of the $Sb_2Te_3$ seed layer.** Large-scale HRTEM images of the PCH thin films with $Sb_2Te_3$ seed layers with a thickness of (a) ~2 nm, (b) ~1 nm and (c) 0 nm. (d)-(f) Atomic-scale HAADF images of the transition region at the interface with the substrate of the three PCH thin films. (g-i) Sketches of the growth scenarios of the three PCH thin films.

We carried out DFT calculations to gain a better understanding of this phenomenon. As discussed before, the Ar-plasma etching of the silicon substrate leads to an ultrathin amorphous silicon slab with dangling bonds and marginal oxygen at the top edge. We constructed hexagonal supercell models of crystalline $TiTe_2$ and $Sb_2Te_3$ and computed the defect formation energy for Si impurities. As shown in Fig. 7, each model contains 60 atoms with 1 Si atom being the substitutional defect, and the cell volume and atomic positions of the models were fully relaxed. We calculated the defect formation energy as $E_d = E_{total} - E_{pristine} - E_{Si} + E_{subst}$, where $E_{total}$, $E_{pristine}$, $E_{Si}$ and $E_{subst}$ represent the energy of the supercell with defect, the pristine phase of the supercell without defect, one Si atom in its bulk ground state, and the one atom being substituted (namely, Ti, Sb or Te atom in the their bulk ground state), respectively. For the substitution in the cation sublattice, the $E_d$ value is 2.48 eV to replace one Ti atom with one Si atom in $TiTe_2$ but 0.79 eV to replace one Sb atom with one Si atom in $Sb_2Te_3$. To replace one Te atom in the anion sublattice, the $E_d$ value is 2.08 eV for $TiTe_2$, and 1.64 eV (edge Te atom) and 1.99 eV (center Te atom) for $Sb_2Te_3$, respectively. Clearly, the defect energies to accommodate Si impurities are smaller in $Sb_2Te_3$ than in $TiTe_2$. We also considered models with Si atoms as interstitial defects, which however yielded much too high energy costs as compared to substitutional ones (Fig. S4) and are ruled out for further discussions. Given the higher substitutional defect energy and the thinner atomic slabs, it is more difficult to form a mixed transition region with $TiTe_2$ than with $Sb_2Te_3$ to generate a smooth surface for the subsequent growth of PCH thin film. These calculations are consistent with our experimental observations shown in Fig. 5 and Fig. 6.

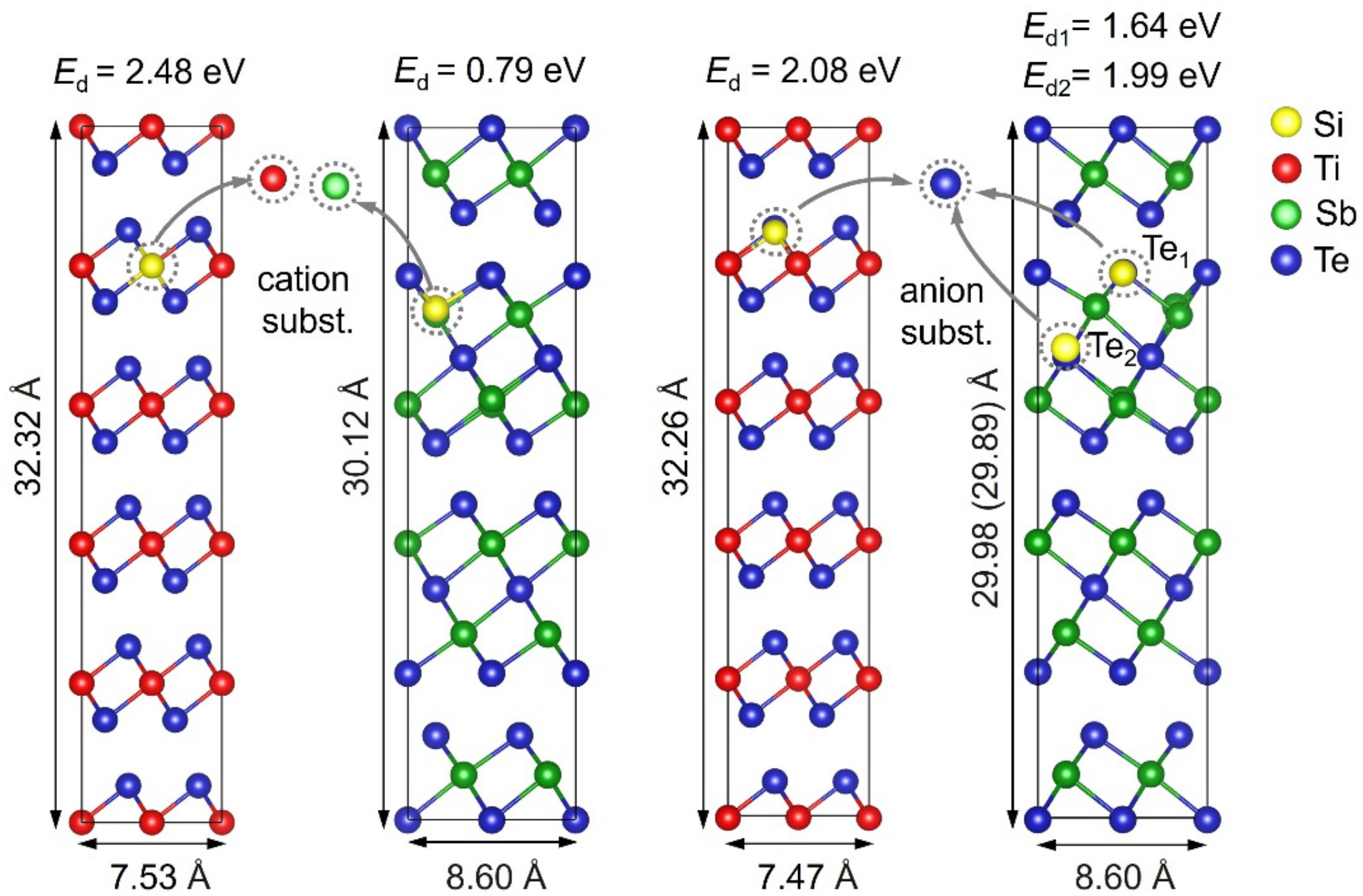

**Fig. 7 Defect energy calculations for Si substitution.** The Si, Ti, Sb, and Te atoms are rendered as yellow, red, green, and blue spheres, respectively. For Te substitution, two non-equivalent sites were considered, denoted as $Te_1$ and $Te_2$.

We also carried out charge transfer analyses based on the DFT-calculated wavefunctions [75], and the obtained Löwdin charge distributions of the five supercell models are displayed in Fig. 8, where the charge values of the Si atom are indicated by the black arrows. In the substituted $TiTe_2$ model, the average charge of Ti atoms is close to 0.6 e, and that of Te atoms is approximately –0.3 e. When the Si atom is incorporated in the Ti sublattice, it tends to gain negative charge rather than to lose it, in contrast to the Ti atoms. This trend becomes more evident for the Te substitution, as the charge of the Si atom reaches –0.6 e. The charge transfer between cation-like Sb and anion-like Te atoms is smaller in $Sb_2Te_3$, and the Si atom acquires negative charge for all the three substitutional sites. In short, the Si impurities act as anion-like atoms rather than cation-like atoms in both $Sb_2Te_3$ and $TiTe_2$, owing to the relatively large electronegativity of Si.

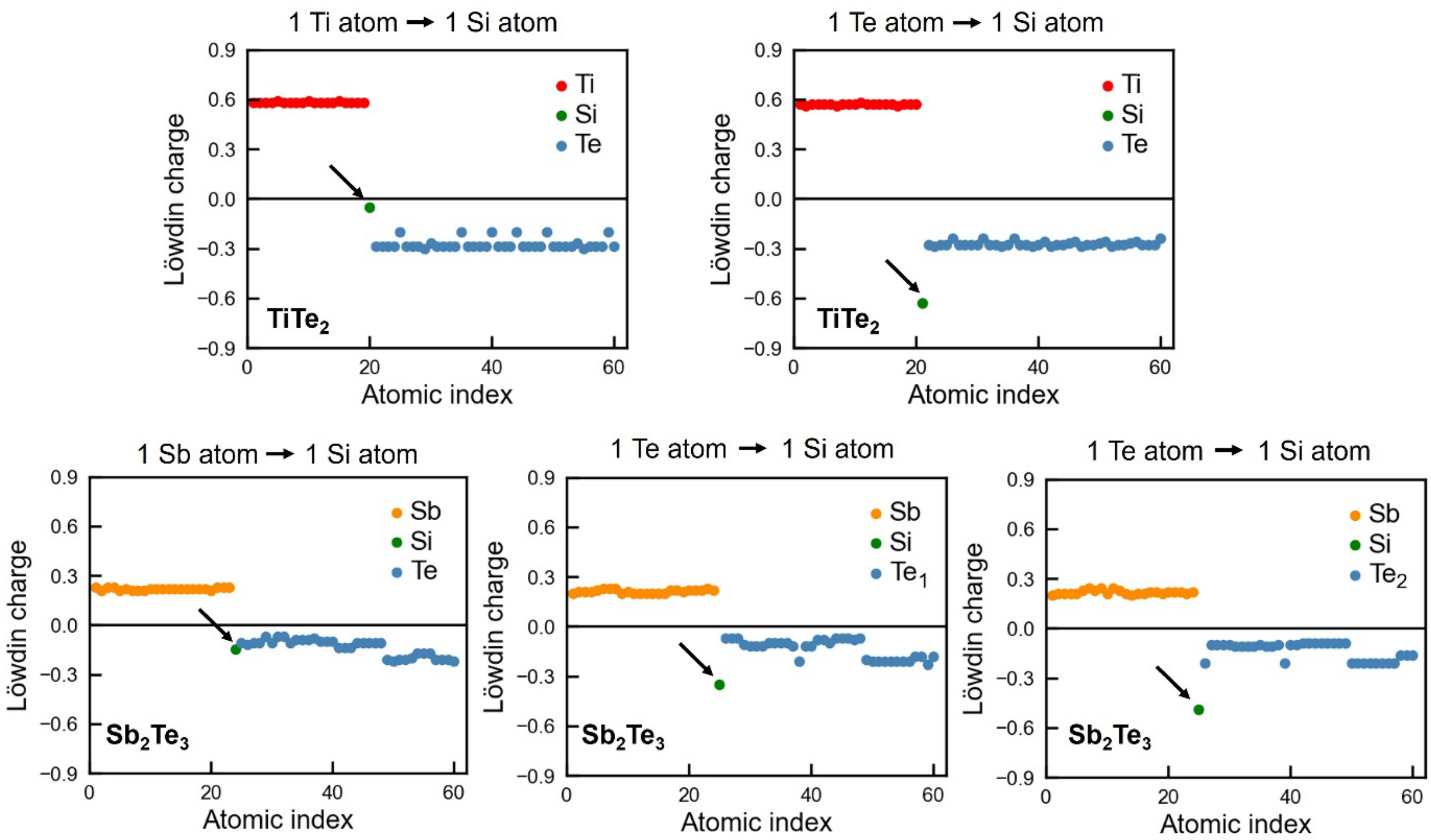

**Fig. 8 Löwdin charge analysis.** The charge values of Si, Ti, Sb, and Te atoms are depicted with green, red, orange, and blue dots, respectively. The charge value of the Si impurity is marked by a black arrow for the five models.

Although it is more frequent to grow $Sb_2Te_3$ as the seed layer for the high-quality growth of $GeTe/Sb_2Te_3$ superlattice thin films [101-103], it could still be possible to generate a flat interface with the silicon substrate when using $TiTe_2$ as the seed layer. In a recent work, an IBM team has reported high-quality $TiTe_2/Sb_2Te_3$ growth in a mushroom device with $TiTe_2$ being the first layer [104], but no details on how to obtain the highly textured seed layer were given. We note that it should be feasible to cover the higher energy costs for mixing $TiTe_2$ with silicon atoms given a higher annealing temperature or a longer annealing time.

After Ar-plasma etching, Noé *et al.* deposited $Sb_2Te_3$ directly on heated silicon substrates, which resulted in Te deficiency in the seed layer [105-107]. The Te-poor seed layer could lead to a disorientation by a few degrees of the atomic planes of the $GeTe/Sb_2Te_3$ superlattice with respect to the substrate [105]. Hence, they proposed to increase the concentration of Te for the seed layer growth, and Te-rich layers were clearly captured by HAADF-EDX measurements [107]. This additional compositional tuning can be avoided by heating the substrate after the deposition of the $Sb_2Te_3$ seed layer is fully completed. Various research teams also reported a complete polishing of the silicon substrates to remove the oxide layers, and epitaxial growth of $Sb_2Te_3$, GST and $GeTe/Sb_2Te_3$ thin films was achieved on top of passivated Si(111) surface with a few Sb atomic layers [108-111] or a single Sb/Te atomic layer [112-114]. Nevertheless, full polishing of the native oxide layer adds fabrication costs, and is not necessary if the procedure sketched in Fig. 1a is adopted.

For our approach, the seed layer needs to be chemically unreactive and intact. Therefore, it should be feasible to use $Bi_2Te_3$, $Bi_2Sb_2Te_5$ and other GST compositions along the GeTe (SnTe) - $Sb_2Te_3$ ($Bi_2Te_3$) pseudo-binary line for the seed layer, as long as a well-ordered trigonal phase can be formed, e.g., $GeSb_4Te_7$, $GeSb_2Te_4$, $Ge_2Sb_2Te_5$, $Ge_3Sb_2Te_6$, $Bi_2Te_3$, $GeBi_2Te_4$, $SnSb_2Te_4$, $SnBi_2Te_4$, etc. [14, 115-126]. However, the compositions between $Ge_8Sb_2Te_{11}$ and GeTe may not be good choices as they only form cubic rocksalt-like structure, which could provide reactive surfaces to interact with $TiTe_2$ layers.

## 4. Conclusion

In conclusion, we discussed in detail how to achieve epitaxial-like growth of $TiTe_2/Sb_2Te_3$ thin film at the wafer-scale, with a particular focus on the essential role of the seed layer. By combining atomic-scale structural characterization and *ab initio* modelling, we revealed the underlying growth mechanism yielding an atomically flat $Sb_2Te_3$ seed layer via a transition slab characterized by a Si-Sb-Te mixture with gradual compositional change. We determined the critical minimum thickness of the seed layer to be ~2 nm with at least one intact ordered $Sb_2Te_3$ QL block at the top edge. Importantly, *in situ* heating should be avoided during the initial Ar-plasma etching and seed layer growth process, since it promotes the re-formation of $SiO_2$ on top of the silicon substrate. However, sufficient post annealing of the seed layer is necessary to form well-ordered $Sb_2Te_3$ QL blocks, prior to the alternate sputtering of the PCH thin film under the *in situ* heating condition. We also observed non-negligible in-plane tensile strain of ~2.4 % in the $TiTe_2$ blocks induced by the upper and lower $Sb_2Te_3$ slabs with larger lattice parameters, which indicates that weak covalent coupling is present across the vdW-like gaps in the PCH thin film. The ultrathin $Sb_2Te_3$ seed layer with critical thickness of ~2 nm could serve as a general starting configuration for highly-textured growth of other chalcogenide thin films at the wafer-scale on a standard silicon substrate with relatively rough $SiO_x$. We also anticipate further exploration of PCH combinations with alternate stacking of transition metal dichalcogenides as confinement layers and main group chalcogenide semiconductors as memory layers, as long as the latter can form chemically unreactive and intrinsically stable atomic blocks.

**Competing interests**

The authors declare no competing interests.

**Data Availability Statement**

Data supporting this work will be available upon requests to the corresponding author.

**Acknowledgments**

The authors thank the support of XJTU for their work at CAID. W.Z. thanks the support of National Natural Science Foundation of China (62374131). J.-J.W. thanks the support of National Natural Science Foundation of China (62204201). The authors acknowledge Chaobin Zeng from Hitachi High-Tech Scientific Solutions (Beijing) Co., Ltd. and Yuanbin Qin for technical support on STEM experiments. The authors also thank Jia Liu at Instrument Analysis Center of XJTU for the assistance regarding Raman spectroscopy experiments. We acknowledge the HPC platform of XJTU and the Computing Center in Xi'an for providing computational resources. The International Joint Laboratory for Micro/Nano Manufacturing and Measurement Technologies of XJTU is acknowledged. R.M. gratefully acknowledges funding from the PRIN 2020 project “Neuromorphic devices based on chalcogenide heterostructures” funded by the Italian Ministry for University and Research (MUR).

**Supplementary materials**

Supplementary material associated with this article can be found in the online version at …

## References

[1] M. Wuttig, N. Yamada, *Nat. Mater.* **2007**, *6*, 824.

[2] W. Zhang, R. Mazzarello, M. Wuttig, E. Ma, *Nat. Rev. Mater.* **2019**, *4*, 150.

[3] A. Sebastian, M. Le Gallo, R. Khaddam-Aljameh, E. Eleftheriou, *Nat. Nanotechnol.* **2020**, *15*, 529.

[4] F. Aguirre, A. Sebastian, M. Le Gallo, W. Song, T. Wang, J. J. Yang, W. Lu, M. F. Chang, D. Ielmini, Y. Yang, A. Mehonic, A. Kenyon, M. A. Villena, J. B. Roldan, Y. Wu, H. H. Hsu, N. Raghavan, J. Sune, E. Miranda, A. Eltawil, G. Setti, K. Smagulova, K. N. Salama, O. Krestinskaya, X. Yan, K. W. Ang, S. Jain, S. Li, O. Alharbi, S. Pazos, M. Lanza, *Nat. Commun.* **2024**, *15*, 1974.

[5] N. Youngblood, C. A. Ríos Ocampo, W. H. P. Pernice, H. Bhaskaran, *Nat. Photon.* **2023**, *17*, 561.

[6] W. Zhou, X. Shen, X. Yang, J. Wang, W. Zhang, *Int. J. Extrem. Manuf.* **2024**, *6*, 022001.

[7] M. Xu, X. Mai, J. Lin, W. Zhang, Y. Li, Y. He, H. Tong, X. Hou, P. Zhou, X. Miao, *Adv. Funct. Mater.* **2020**, *30*, 2003419.

[8] X. D. Li, N. K. Chen, B. Q. Wang, M. Niu, M. Xu, X. Miao, X.-B. Li, *Adv. Mater.* **2024**, *36*, 2307951.

[9] B. Liu, K. Li, J. Zhou, Z. Sun, *Adv. Funct. Mater.* **2024**, *34*, 2407239.

[10] X. Wang, R. Wang, S. Sun, D. Xu, C. Nie, Z. Zhou, C. Wen, J. Zhang, R. Chu, X. Shen, W. Zhou, Z. Song, J. Wang, E. Ma, W. Zhang, *arXiv: 2503.21446* **2025**.

[11] X. Shen, R. Chu, Y. Jiang, W. Zhang, *Acta Metall. Sin.* **2024**, *60*, 1362.

[12] X. Shen, S. Zhang, Y. Jiang, T. Huang, S. Sun, W. Zhou, J. Wang, R. Mazzarello, W. Zhang, *MGE Adv.* **2024**, *2*, e62.

[13] N. Yamada, E. Ohno, N. Akahira, K. Nishiuchi, K. Nagata, M. Takao, *Jpn. J. Appl. Phys. Part 1* **1987**, *26*, 61.

[14] N. Yamada, E. Ohno, K. Nishiuchi, N. Akahira, M. Takao, *J. Appl. Phys.* **1991**, *69*, 2849.

[15] I. Friedrich, V. Weidenhof, W. Njoroge, P. Franz, M. Wuttig, *J. Appl. Phys.* **2000**, *87*, 4130.

[16] B. J. Kooi, J. T. M. De Hosson, *J. Appl. Phys.* **2002**, *92*, 3584.

[17] B. Zhang, W. Zhang, Z.-J. Shen, Y.-J. Chen, J.-X. Li, S.-B. Zhang, Z. Zhang, M. Wuttig, R. Mazzarello, E. Ma, X.-D. Han, *Appl. Phys. Lett.* **2016**, *108*, 191902.

[18] T. Siegrist, P. Jost, H. Volker, M. Woda, P. Merkelbach, C. Schlockermann, M. Wuttig, *Nat. Mater.* **2011**, *10*, 202.

[19] W. Zhang, M. Wuttig, R. Mazzarello, *Sci. Rep.* **2015**, *5*, 13496.

[20] T.-T. Jiang, X.-D. Wang, J.-J. Wang, Y.-X. Zhou, D.-L. Zhang, L. Lu, C.-L. Jia, M. Wuttig, R. Mazzarello, W. Zhang, *Acta Mater.* **2020**, *187*, 103.

[21] A. Lotnyk, M. Behrens, B. Rauschenbach, *Nanoscale Adv.* **2019**, *1*, 3836.

[22] H.-S. P. Wong, S. Raoux, S. B. Kim, J. Liang, J. P. Reifenberg, B. Rajendran, M. Asheghi, K. E. Goodson, *Proc. IEEE* **2010**, *98*, 2201.

[23] X. Li, H. Chen, C. Xie, D. Cai, S. Song, Y. Chen, Y. Lei, M. Zhu, Z. Song, *Phys. Status Solidi RRL* **2019**, *13*, 1800558.

[24] H. Y. Cheng, F. Carta, W. C. Chien, H. L. Lung, M. BrightSky, *J. Phys. D: Appl. Phys.* **2019**, *52*, 473002

[25] P. Cappelletti, R. Annunziata, F. Arnaud, F. Disegni, A. Maurelli, P. Zuliani, *J. Phys. D: Appl. Phys.* **2020**, *53*, 193002.

[26] Q. Wang, Y. Wang, Y. Wang, L. Jiang, J. Zhao, Z. Song, J. Bi, L. Zhao, Z. Jiang, J. Schwarzkopf, S. Wu, B. Zhang, W. Ren, S. Song, G. Niu, *InfoMat* **2024**, *6*, e12543.

[27] C. Ríos, M. Stegmaier, P. Hosseini, D. Wang, T. Scherer, C. D. Wright, H. Bhaskaran, W. H. P. Pernice, *Nat. Photon.* **2015**, *9*, 725.

[28] P. Hosseini, C. D. Wright, H. Bhaskaran, *Nature* **2014**, *511*, 206.

[29] W. Zhou, B. Dong, N. Farmakidis, X. Li, N. Youngblood, K. Huang, Y. He, C. David Wright, W. H. P. Pernice, H. Bhaskaran, *Nat. Commun.* **2023**, *14*, 2887.

[30] W. Zhang, R. Mazzarello, E. Ma, *MRS Bull.* **2019**, *44*, 686.

[31] D. Ielmini, A. L. Lacaita, D. Mantegazza, *IEEE Trans. Electron. Dev.* **2007**, *54*, 308.

[32] S. Braga, A. Cabrini, G. Torelli, *Applied Physics Letters* **2009**, *94*, 092112.

[33] M. Boniardi, D. Ielmini, S. Lavizzari, A. L. Lacaita, A. Redaelli, A. Pirovano, *IEEE Trans. Electron. Dev.* **2010**, *57*, 2690.

[34] M. Boniardi, D. Ielmini, *Appl. Phys. Lett.* **2011**, *98*, 243506.

[35] A. Sebastian, M. Le Gallo, G. W. Burr, S. Kim, M. BrightSky, E. Eleftheriou, *J. Appl. Phys.* **2018**, *124*, 111101.
[36] M. Le Gallo, A. Sebastian, R. Mathis, M. Manica, H. Giefers, T. Tuma, C. Bekas, A. Curioni, E. Eleftheriou, *Nat. Electron.* **2018**, *1*, 246.
[37] S. Ambrogio, P. Narayanan, H. Tsai, R. M. Shelby, I. Boybat, G. W. Burr, *Nature* **2018**, *558*, 60.
[38] J.-Y. Raty, W. Zhang, J. Luckas, C. Chen, C. Bichara, R. Mazzarello, M. Wuttig, *Nat. Commun.* **2015**, *6*, 7467.
[39] S. Gabardi, S. Caravati, G. C. Sosso, J. Behler, M. Bernasconi, *Phys. Rev. B* **2015**, *92*, 054201.
[40] K. Konstantinou, F. C. Mocanu, T. H. Lee, S. R. Elliott, *Nat. Commun.* **2019**, *10*, 3065.
[41] W. Zhang, E. Ma, *Mater. Today* **2020**, *41*, 156.
[42] B. Chen, X. P. Wang, F. Jiao, L. Ning, J. Huang, J. Xie, S. Zhang, X. B. Li, F. Rao, *Adv. Sci.* **2023**, *10*, 2301043.
[43] F. Jiao, B. Chen, K. Ding, K. Li, L. Wang, X. Zeng, F. Rao, *Appl. Mater. Today* **2020**, *20*, 100641.
[44] P. Ma, H. Tong, M. Xu, X. Cheng, X. Miao, *Appl. Phys. Lett.* **2020**, *117*, 022109
[45] X. Shen, Y. Zhou, H. Zhang, V. L. Deringer, R. Mazzarello, W. Zhang, *Nanoscale* **2023**, *15*, 15259.
[46] J. Huang, B. Chen, G. Sha, H. Gong, T. Song, K. Ding, F. Rao, *Nano Lett.* **2023**, *23*, 2362.
[47] B. Liu, K. Li, W. Liu, J. Zhou, L. Wu, Z. Song, S. R. Elliott, Z. Sun, *Sci. Bull.* **2021**, *66*, 2217.
[48] Y. Jiao, G. Wang, A. Lotnyk, T. Wu, J. Zhu, A. He, *J. Alloys Compd.* **2023**, *953*, 169970.
[49] A. He, J. Zhu, G. Wang, A. Lotnyk, S. Cremer, Y. Chen, X. Shen, *Appl. Phys. Lett.* **2024**, *124*, 221602.
[50] T. C. Chong, L. P. Shi, R. Zhao, P. K. Tan, J. M. Li, H. K. Lee, X. S. Miao, A. Y. Du, C. H. Tung, *Appl. Phys. Lett.* **2006**, *88*, 122114.
[51] R. E. Simpson, P. Fons, A. V. Kolobov, T. Fukaya, M. Krbal, T. Yagi, J. Tominaga, *Nat. Nanotechnol.* **2011**, *6*, 501.
[52] R. Wang, V. Bragaglia, J. E. Boschker, R. Calarco, *Cryst. Growth Des.* **2016**, *16*, 3596.
[53] A. V. Kolobov, P. Fons, Y. Saito, J. Tominaga, *ACS Omega* **2017**, *2*, 6223.
[54] A. Lotnyk, I. hilmi, U. Ross, B. Rauschenbach, *Nano Res.* **2017**, *11*, 1676.
[55] L. Zhou, Z. Yang, X. Wang, H. Qian, M. Xu, X. Cheng, H. Tong, X. Miao, *Adv. Electron. Mater.* **2019**, *5*, 1900781.
[56] C. Yoo, J. W. Jeon, B. Park, W. Choi, G. Jeon, S. Jeon, S. Kim, C. S. Hwang, *ACS Appl. Electron. Mater.* **2023**, *5*, 5794.
[57] X.-B. Li, N.-K. Chen, X.-P. Wang, H.-B. Sun, *Adv. Funct. Mater.* **2018**, *28*, 1803380.
[58] B. Sa, J. Zhou, Z. Sun, J. Tominaga, R. Ahuja, *Phys. Rev. Lett.* **2012**, *109*, 096802.
[59] O. Cojocaru-Mirédin, J.-C. Bürger, N. Polin, A. Meledin, J. Mayer, M. Wuttig, A. Daus, *ACS nano* **2025**, , DOI: 10.1021/acsnano.4c13450.
[60] K. Ding, J. Wang, Y. Zhou, H. Tian, L. Lu, R. Mazzarello, C. Jia, W. Zhang, F. Rao, E. Ma, *Science* **2019**, *366*, 210.
[61] J. Shen, S. Lv, X. Chen, T. Li, S. Zhang, Z. Song, M. Zhu, *ACS Appl. Mater. Interfaces* **2019**, *11*, 5336.
[62] X. Wang, K. Ding, M. Shi, J. Li, B. Chen, M. Xia, J. Liu, Y. Wang, J. Li, E. Ma, Z. Zhang, H. Tian, F. Rao, *Mater. Today* **2022**, *54*, 52.
[63] X. Wang, Y. Wu, Y. Zhou, V. L. Deringer, W. Zhang, *Mater. Sci. Semicond. Process.* **2021**, *135*, 106080.
[64] G. Han, F. Liu, Y. Zhang, J. Li, W. Li, Q. Chen, Y. Li, X. Xie, *J. Alloys Compd.* **2021**, *877*, 160270.
[65] J. Park, H. Sung, S. Son, S. Ju, H. Lee, *Phys. Status Solidi RRL* **2023**, *17*, 2200451.
[66] R. Piombo, S. Ritarossi, R. Mazzarello, *Adv. Sci.* **2024**, *11*, 2402375.
[67] T. H. Kim, S. W. Park, H. J. Lee, D. H. Kim, J. Y. Choi, T. G. Kim, *Small* **2023**, *19*, 2303659.
[68] T. H. Kim, K. J. Yoo, T. H. Kim, H. J. Lee, A. C. Khot, K. A. Nirmal, S. H. Hong, T. G. Kim, *Appl. Surf. Sci.* **2023**, *626*, 157291.
[69] S. W. Park, H. J. Lee, K. A. Nirmal, T. H. Kim, D. H. Kim, J. Y. Choi, J. S. Oh, J. M. Joo, T. G. Kim, *J. Mater. Sci. Technol.* **2025**, *204*, 104.
[70] X. Wu, A. I. Khan, H. Lee, C.-F. Hsu, H. Zhang, H. Yu, N. Roy, A. V. Davydov, I. Takeuchi, X. Bao, H. S. P. Wong, E. Pop, *Nat. Commun.* **2024**, *15*, 13.
[71] G. Kresse, J. Hafner, *Phys. Rev. B* **1993**, *47*, 558.
[72] J. P. Perdew, K. Burke, M. Ernzerhof, *Phys. Rev. Lett.* **1996**, *77*, 3865.

[73] G. Kresse, D. Joubert, *Phys. Rev. B* **1999**, *59*, 1758.
[74] S. Grimme, J. Antony, S. Ehrlich, H. Krieg, *J. Chem. Phys.* **2010**, *132*, 154104
[75] R. Nelson, C. Ertural, J. George, V. L. Deringer, G. Hautier, R. Dronskowski, *J. Comput. Chem.* **2020**, *41*, 1931.
[76] S. Maintz, V. L. Deringer, A. L. Tchougreeff, R. Dronskowski, *J. Comput. Chem.* **2016**, *37*, 1030.
[77] V. L. Deringer, A. L. Tchougreeff, R. Dronskowski, *J. Phys. Chem. A.* **2011**, *115*, 5461.
[78] Y. Saito, P. Fons, A. V. Kolobov, J. Tominaga, *Phys. Status Solidi B* **2015**, *252*, 2151.
[79] Y. Saito, P. Fons, L. Bolotov, N. Miyata, A. V. Kolobov, J. Tominaga, *AIP Adv.* **2016**, *6*, 045220.
[80] Y. Saito, P. Fons, A. V. Kolobov, K. V. Mitrofanov, K. Makino, J. Tominaga, S. Hatayama, Y. Sutou, M. Hase, J. Robertson, *J. Phys. D: Appl. Phys.* **2020**, *53*, 284002.
[81] T. L. Anderson, H. B. Krause, *Acta Crystallogr. B* **1974** *30* 1307.
[82] Y. Zheng, M. Xia, Y. Cheng, F. Rao, K. Ding, W. Liu, Y. Jia, Z. Song, S. Feng, *Nano Res.* **2016**, *9*, 3453.
[83] Y. Xu, X. Wang, W. Zhang, L. Schäfer, J. Reindl, F. vom Bruch, Y. Zhou, V. Evang, J. J. Wang, V. L. Deringer, E. Ma, M. Wuttig, R. Mazzarello, *Adv. Mater.* **2021**, *33*, 2006221.
[84] W. Richter, C. R. Becker, *Phys. Status Solidi B* **1977**, *84*, 619.
[85] K. M. F. Shahil, M. Z. Hossain, V. Goyal, A. A. Balandin, *J. Appl. Phys.* **2012**, *111*, 054305.
[86] G. C. Sosso, S. Caravati, M. Bernasconi, *J. Phys. Condens. Matter.* **2009**, *21*, 095410.
[87] X. Feng, Z. Li, G. Chen, H. Yue, Y. Gao, X. Zhang, Z. Guo, W. Yuan, *CrystEngComm* **2023**, *25*, 5399.
[88] S. J. Pennycook, P. D. Nellist, *Scanning Transmission Electron Microscopy Imaging and Analysis*, Springer, 2011.
[89] R. Wang, F. R. L. Lange, S. Cecchi, M. Hanke, M. Wuttig, R. Calarco, *Adv. Funct. Mater.* **2018**, *28*, 1705901.
[90] J. Wang, I. Ronneberger, L. Zhou, L. Lu, V. L. Deringer, B. Zhang, L. Tian, H. Du, C. Jia, X. Qian, M. Wuttig, R. Mazzarello, W. Zhang, *Nanoscale* **2018**, *10*, 7363.
[91] Y. Cheng, O. Cojocaru-Mirédin, J. Keutgen, Y. Yu, M. Küpers, M. Schumacher, P. Golub, J.-Y. Raty, R. Dronskowski, M. Wuttig, *Adv. Mater.* **2019**, *31*, 1904316.
[92] J. Y. Raty, M. Schumacher, P. Golub, V. L. Deringer, C. Gatti, M. Wuttig, *Adv. Mater.* **2019**, *31*, 1806280.
[93] B. J. Kooi, M. Wuttig, *Adv. Mater.* **2020**, *32*, 1908302.
[94] M. Wuttig, C. F. Schon, J. Lotfering, P. Golub, C. Gatti, J. Y. Raty, *Adv. Mater.* **2023**, *35*, 2208485.
[95] M. Wuttig, C. F. Schon, D. Kim, P. Golub, C. Gatti, J. Y. Raty, B. J. Kooi, A. M. Pendas, R. Arora, U. Waghmare, *Adv. Sci.* **2023**, *11*, 2308578.
[96] W. Zhang, H. Zhang, S. Sun, X. Wang, Z. Lu, X. Wang, J.-J. Wang, C. Jia, C. F. Schon, R. Mazzarello, E. Ma, M. Wuttig, *Adv. Sci.* **2023**, *10*, 2300901.
[97] M. Chhowalla, H. S. Shin, G. Eda, L. J. Li, K. P. Loh, H. Zhang, *Nat. Chem.* **2013**, *5*, 263.
[98] J. Qi, X. Qian, L. Qi, J. Feng, D. Shi, J. Li, *Nano Lett.* **2012**, *12*, 1224.
[99] J. Li, Z. Shan, E. Ma, *MRS Bull.* **2014**, *39*, 108.
[100] W. Li, X. Qian, J. Li, *Nat. Rev. Mater.* **2021**, *6*, 829.
[101] J. Momand, F. R. L. Lange, R. Wang, J. E. Boschker, M. A. Verheijen, R. Calarco, M. Wuttig, B. J. Kooi, *J. Mater. Res.* **2016**, *31*, 3115.
[102] A. I. Khan, A. Daus, R. Islam, K. M. Neilson, H. R. Lee, H.-S. P. Wong, E. Pop, *Science* **2021**, *373*, 1243.
[103] S. Prili, V. Bragaglia, V. P. Jonnalagadda, J. Luchtenveld, B. J. Kooi, F. Arciprete, A. Sebastian, G. S. Syed, *arXiv: 2411.19343* **2024**.
[104] G. M. Cohen, A. Majumdar, C. W. Cheng, A. Ray, D. Piatek, L. Gignac, C. Lavoie, A. Grun, H. Y. Cheng, Z. L. Liu, H. L. Lung, H. Miyazoe, R. L. Bruce, M. BrightSky, *Phys. Status Solidi RRL* **2024**, *18*, 2300426.
[105] P. Kowalczyk, F. Hippert, N. Bernier, C. Mocuta, C. Sabbione, W. Batista‐Pessoa, P. Noé, *Small* **2018**, *14*, 1704514.
[106] F. Hippert, P. Kowalczyk, N. Bernier, C. Sabbione, X. Zucchi, D. Térébénec, C. Mocuta, P. Noé, *J. Phys. D: Appl.*

*Phys.* **2020**, *53*, 154003.
[107] V. Sever, N. Bernier, D. Térébénec, C. Sabbione, J. Paterson, F. Castioni, P. Quéméré, A. Jannaud, J. L. Rouvière, H. Roussel, J. Y. Raty, F. Hippert, P. Noé, *Phys. Status Solidi RRL* **2024**, *18*, 2300402.
[108] J. Momand, R. Wang, J. E. Boschker, M. A. Verheijen, R. Calarco, B. J. Kooi, *Nanoscale* **2015**, *7*, 19136.
[109] R. Wang, R. Calarco, F. Arciprete, V. Bragaglia, *Mater. Sci. Semicond. Process.* **2022**, *137*, 106244.
[110] V. Bragaglia, F. Arciprete, W. Zhang, A. M. Mio, E. Zallo, K. Perumal, A. Giussani, S. Cecchi, J. E. Boschker, H. Riechert, S. Privitera, E. Rimini, R. Mazzarello, R. Calarco, *Sci. Rep.* **2016**, *6*, 23843.
[111] V. Bragaglia, F. Arciprete, A. M. Mio, R. Calarco, *J. Appl. Phys.* **2018**, *123*.
[112] I. Hilmi, A. Lotnyk, J. W. Gerlach, P. Schumacher, B. Rauschenbach, *APL Materials* **2017**, *5*, 050701.
[113] M. Behrens, A. Lotnyk, U. Roß, J. Griebel, P. Schumacher, J. W. Gerlach, B. Rauschenbach, *CrystEngComm* **2018**, *20*, 3688.
[114] H. Bryja, J. W. Gerlach, A. Prager, M. Ehrhardt, B. Rauschenbach, A. Lotnyk, *2D Materials* **2021**, *8*, 045027.
[115] J.-W. Park, S. H. Eom, H. Lee, J. L. F. Da Silva, Y.-S. Kang, T.-Y. Lee, Y. H. Khang, *Phys. Rev. B* **2009**, *80*, 115209.
[116] J. Da Silva, A. Walsh, H. Lee, *Phys. Rev. B* **2008**, *78*, 224111.
[117] Z. Sun, Y. Pan, J. Zhou, B. Sa, R. Ahuja, *Phys. Rev. B* **2011**, *83*, 113201.
[118] Y. Gan, J. Zhou, Z. Sun, *J. Mater. Inf.* **2021**, *1*, 2.
[119] Y. Gan, Y. Huang, N. Miao, J. Zhou, Z. Sun, *J. Mater. Chem. C* **2021**, *9*, 4189.
[120] J. J. Wang, H. M. Zhang, X. D. Wang, L. Lu, C. Jia, W. Zhang, R. Mazzarello, *Adv. Mater. Technol.* **2022**, *7*, 2200214.
[121] J.-J. Wang, J. Wang, H. Du, L. Lu, P. C. Schmitz, Johannes Reindl, A. M. Mio, C.-L. Jia, E. Ma, R. Mazzarello, M. Wuttig, W. Zhang, *Chem. Mater.* **2018**, *30*, 4770.
[122] H. Zhang, D. T. Yimam, S. de Graaf, J. Momand, P. A. Vermeulen, Y. Wei, B. Noheda, B. J. Kooi, *ACS nano* **2021**, *15*, 2869.
[123] H. Zhang, J. Momand, J. Levinsky, Q. Guo, X. Zhu, G. H. ten Brink, G. R. Blake, G. Palasantzas, B. J. Kooi, *Nano Res.* **2021**, *15*, 2382.
[124] D. Hsieh, Y. Xia, D. Qian, L. Wray, F. Meier, J. Dil, J. Osterwalder, L. Patthey, A. Fedorov, H. Lin, A. Bansil, D. Grauer, Y. Hor, R. Cava, M. Hasan, *Phys. Rev. Lett.* **2009**, *103*, 146401.
[125] S. Howard, A. Raghavan, D. Iaia, C. Xu, D. Flötotto, M.-H. Wong, S.-K. Mo, B. Singh, R. Sankar, H. Lin, T.-C. Chiang, V. Madhavan, *Phys. Rev. Mater.* **2022**, *6*, 044201
[126] T. Schäfer, P. M. Konze, J. D. Huyeng, V. L. Deringer, T. Lesieur, P. Müller, M. Morgenstern, R. Dronskowski, M. Wuttig, *Chem. Mater.* **2017**, *29*, 6749.

# Supporting Information

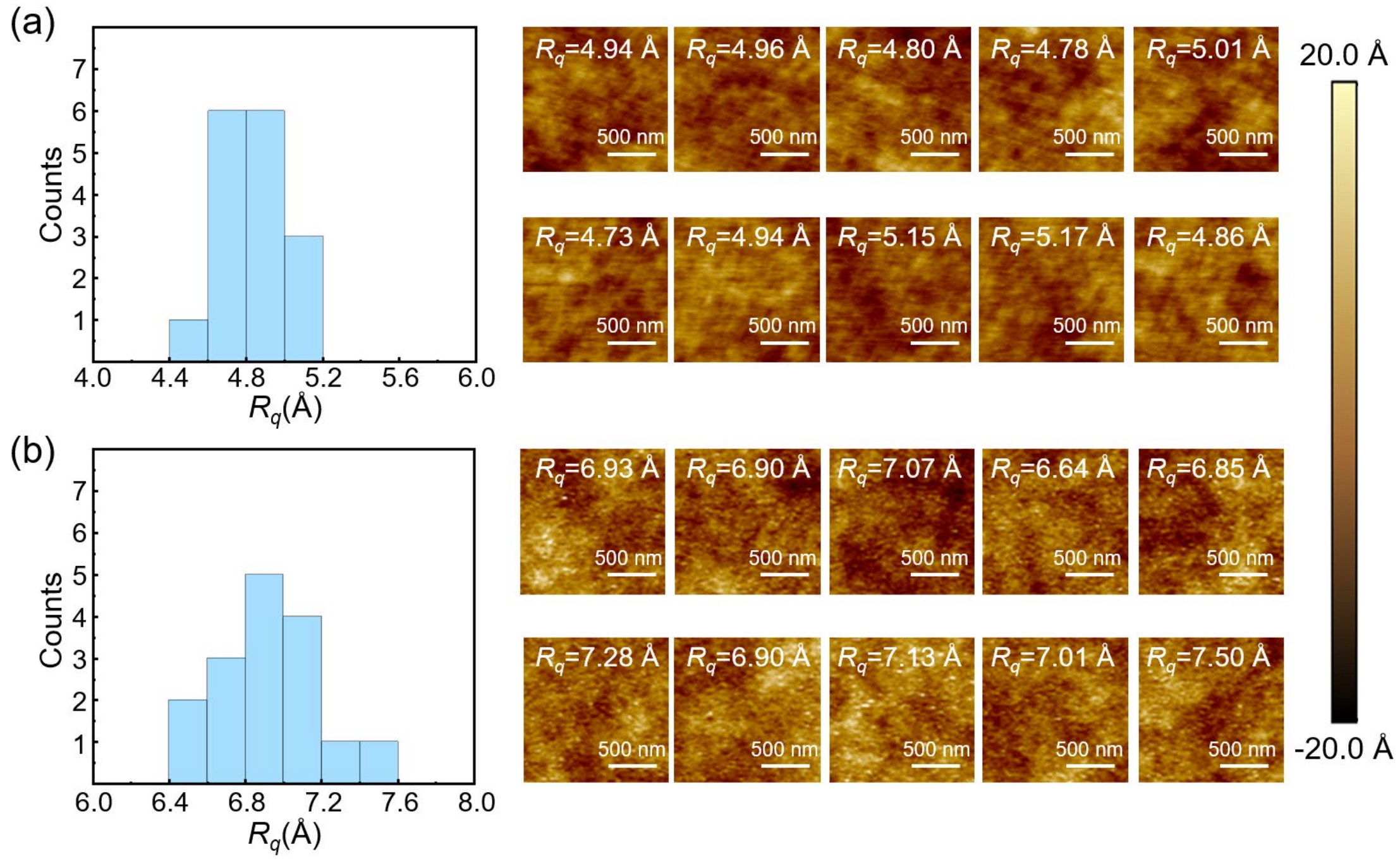

**Fig. S1** Statistical histogram of the root-mean-square roughness $R_q$ for 16 different regions, and the AFM images for ten of them. (a) ~5nm thick $Sb_2Te_3$ seed layer and (b) ~45nm thick $Sb_2Te_3/TiTe_2$ film. The $R_q$ of $Sb_2Te_3$ seed layer and $Sb_2Te_3/TiTe_2$ film are observed to be below 5.2 Å and 7.6 Å, respectively.

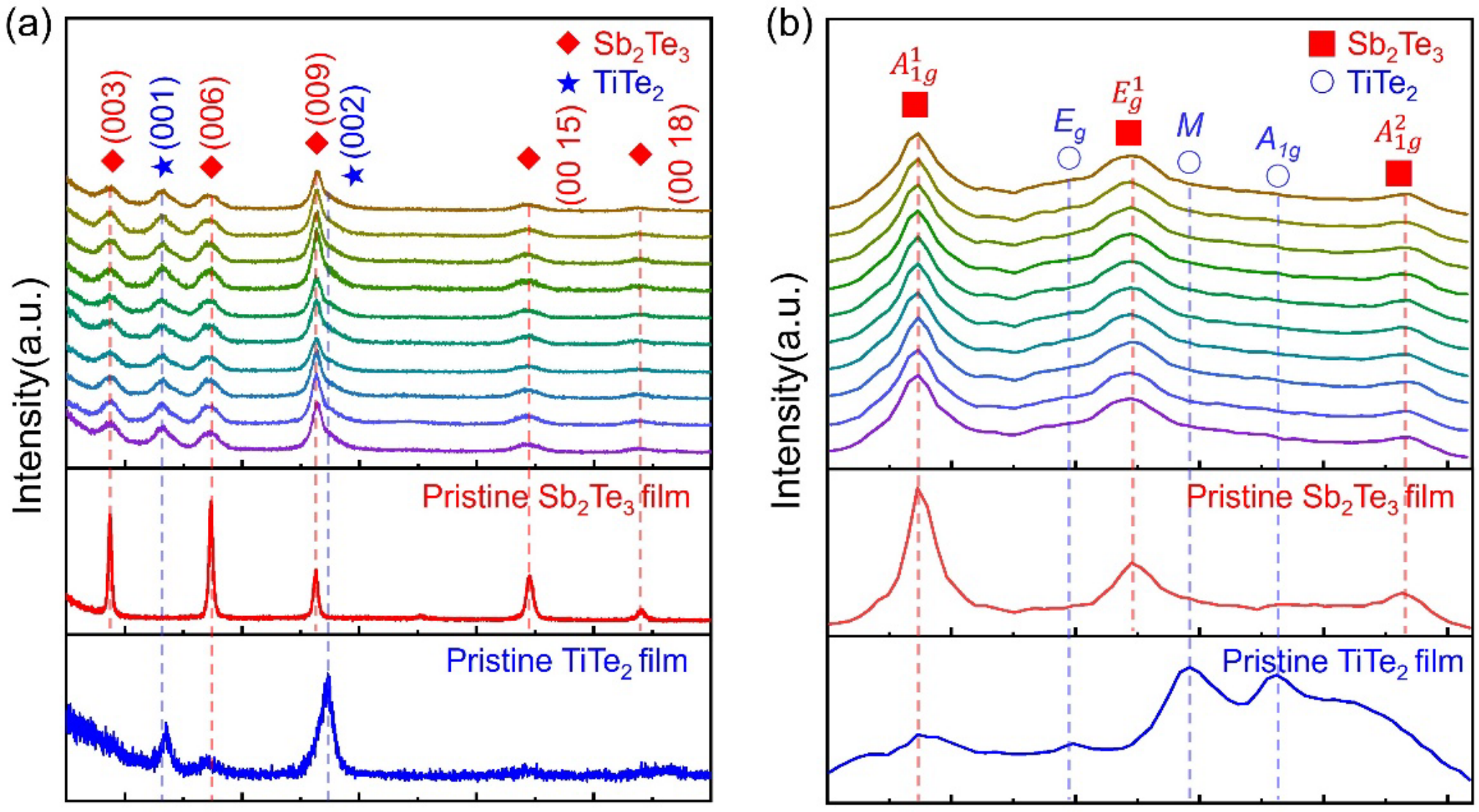

**Fig. S2** (a) XRD patterns and (b) Raman spectra for ten different sampling spots in $Sb_2Te_3/TiTe_2$ thin films.

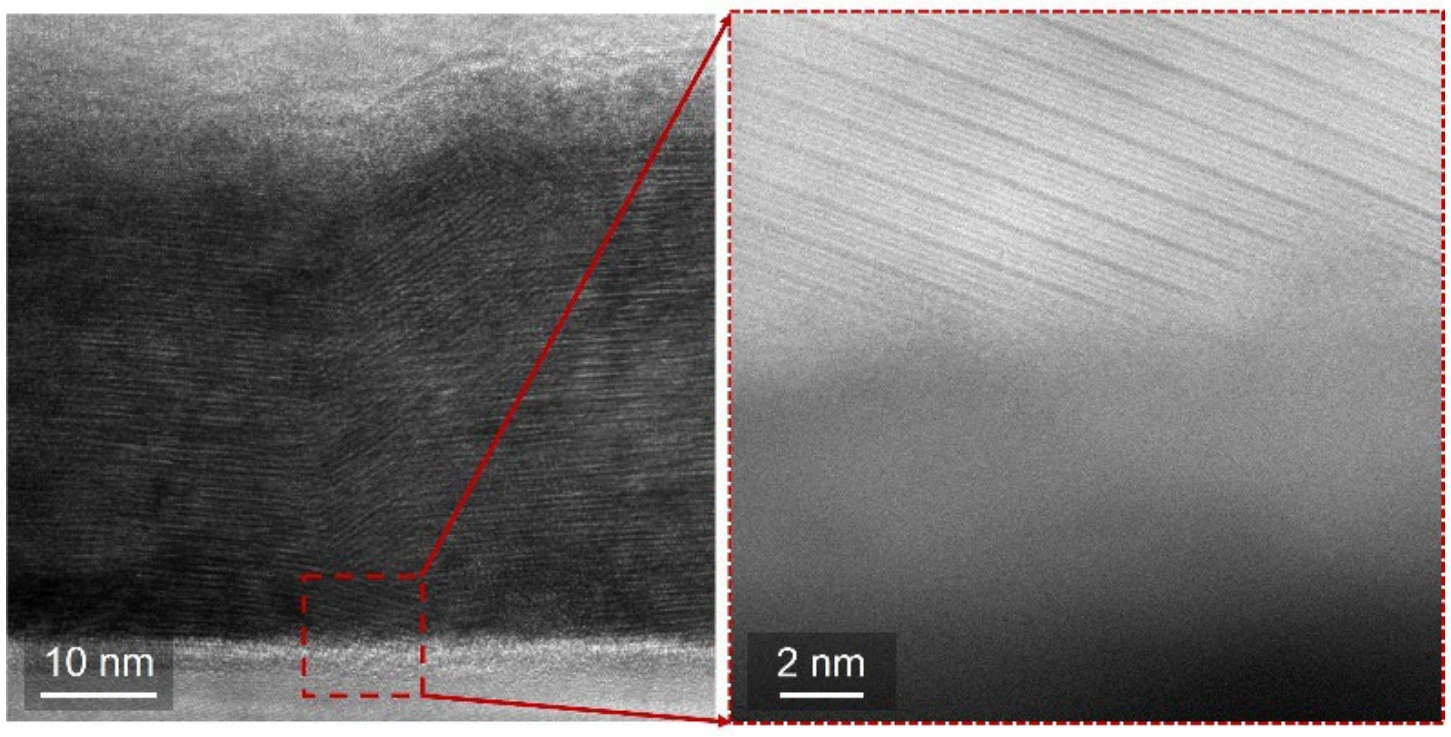

**Fig. S3** The TEM characterizations of the $TiTe_2/Sb_2Te_3$ thin film, for which the growth of the $Sb_2Te_3$ seed layer was also subjected to in situ heating condition.

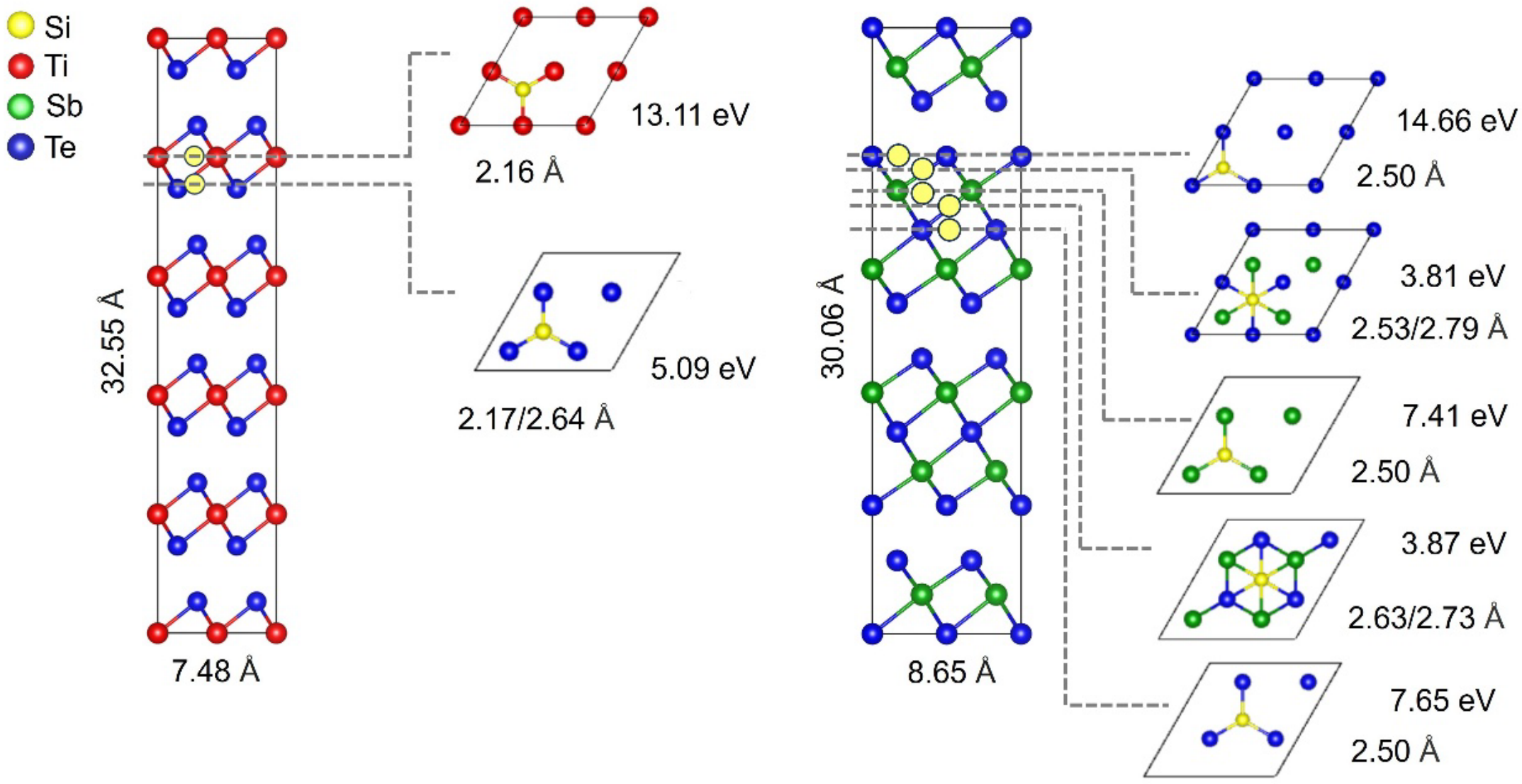

**Fig. S4 Defect formation energy ($E_d$) of Si interstitial sites.** The $E_d$ value is calculated as $E_d = E_{total} - E_{pristine} - E_{Si}$, where $E_{total}$, $E_{pristine}$ and $E_{Si}$ represent the energy of the supercell with one interstitial Si atom, the pristine phase of the supercell, and the energy of one Si atom in its bulk ground state. The Si, Ti, Sb, and Te atoms are rendered as yellow, red, green, and blue spheres, respectively. The Si interstitial atom adds much higher energy costs, and would induce large atomic displacement of the surrounding atoms upon relaxation.